\documentclass[9pt,twocolumn,twoside]{osajnl}

\journal{ao} 

\setboolean{shortarticle}{false}

\usepackage[justification=justified]{caption}
\usepackage[justification=centering, margin=5pt]{subcaption}
\graphicspath{{./Figures/}}
\usepackage{lscape}

\usepackage{dblfloatfix}
\usepackage{multirow}

\definecolor{Green}{rgb}{0, 0.5, 0}
\definecolor{Aqua}{rgb}{0.35, 0.7, 0.7}

\usepackage[symbol]{footmisc}

\title{A Step-by-Step Guide to 3D Print Motorized Rotation Mounts for Optical Applications}

\author[ ]{Daniel P.G. Nilsson}
\author[ ]{Tobias Dahlberg}
\author[*]{Magnus Andersson}

\affil[ ]{Department of Physics, Umeå University, 901 87 Umeå, Sweden}

\affil[*]{Corresponding author: magnus.andersson@umu.se}

\doi{\url{}}

\begin{abstract}
Motorized rotation mounts and stages are versatile instruments that introduce computer control to optical systems, enabling automation and scanning actions. They can be used for intensity control and position adjustments, etc. However, these rotation mounts come with a hefty price tag, and this limits their use. This work shows how to build two different types of motorized rotation mounts for 1" optics, using a 3D printer and off-the-shelf components. The first is intended for reflective elements, like mirrors and gratings, and the second for transmissive elements, like polarizers and retarders. We evaluate and compare their performance to commercial systems based on velocity, resolution, accuracy, backlash, and axis wobble. Also, we investigate the angular stability using Allan variance analysis. The results show that our mounts perform similar to systems costing more than \texteuro 2000, while also being quick to build and costing less than \texteuro 200. As a proof of concept, we show how to control lasers used in an optical tweezers and Raman spectroscopy setup. When used for this, the 3D printed motorized rotational mounts provide intensity control with a resolution of $0.03$ percentage points or better.

\end{abstract}

\setboolean{displaycopyright}{true}

\begin{document}

\maketitle

\section{Introduction}
\label{sec:Introduction}
Rotation mounts and rotary stages are essential in many optical systems. They are versatile and can be used to adjust the polarization direction in an attenuator setup, rotate the grating in a spectrometer, or change the position of mirrors and filters, among other things \cite{bib:Lotem1991, bib:Fueten1997}. Having these optical elements motorized and controlled by a computer increases their usability by allowing higher accuracy and repeatability, as well as the possibility for automation and scanning actions. Unfortunately, commercial rotation mounts and their corresponding drivers come with a hefty price tag, discouraging their use.

On the contrary, 3D printers nowadays come with a fair price tag thanks to the Do-It-Yourself community \cite{bib:Tully2020} and they have gained significance in experimental research groups to develop laboratory-specific equipment \cite{baden2015open,winters20183d}. This development has in turn made key components of both 3D printers and rotation mounts, like stepper motors and drivers, inexpensive and readily available. Based on this, would it be possible to design rotation mounts that combine the advantages of 3D printing with the accessibility of its components? If so, how would its performance and cost compare to similar systems made by industry-leading manufacturers? 

Previous papers addressing in-house rotation mounts \cite{bib:Shelton2011, bib:Rakonjac2013} use fabrication techniques outside of the capabilities of the common lab. They developed rotation mounts that were either unsuitable for standard optics or used components that are not feasible to produce a low cost product. They also used a piezoelectric rotor, which are known to cause vibrations, that can disturb sensitive measurements. Therefore, in this work, we design multipurpose rotation mounts for applications that uses 1" optics ($1\,in. = 2.54\,cm$). We build the mounts using a commercial 3D printer [\textit{Ultimaker 2+}], then we test and evaluate their performance. We constraint the budget to \texteuro 200 per mount (incl. driver) and provide in-depth instructions and examples for building and integrating them into an optical setup.

\section{Materials And Methods}
\label{sec:MaterialsAndMethods}

\subsection{3D Printing Rotational Mounts}
\label{subsec:RotationMount}
\begin{figure}[!t]
    \centering
    \includegraphics[width=\linewidth]{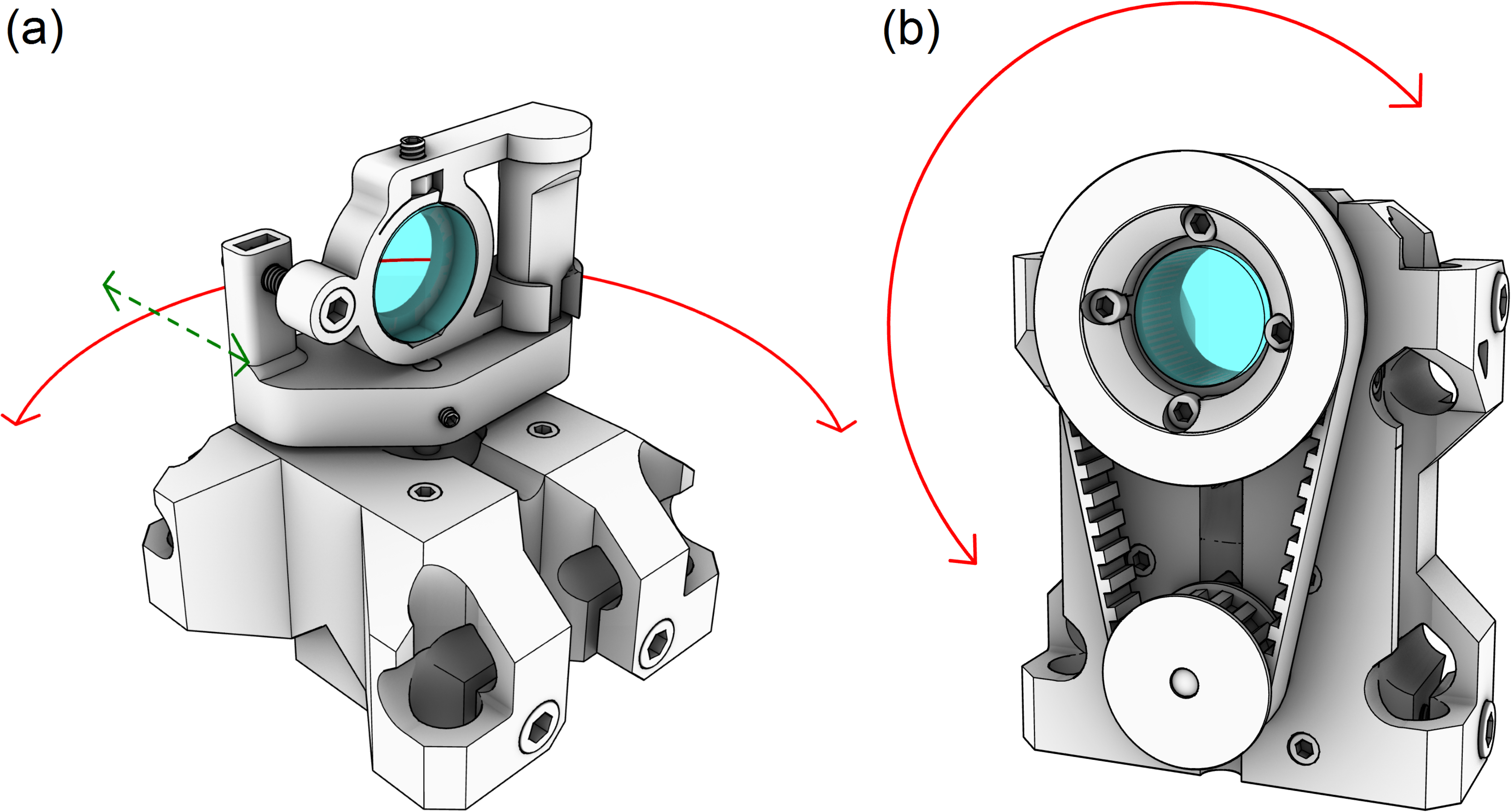}
    \caption{The reflective mount (a) and transmissive mount (b) are 3D printed rotation mounts for different optical \textcolor{Aqua}{elements}. Arrows show the rotation (\textcolor{red}{solid}) and position adjustment (\textcolor{Green}{dashed}) for the mounts, which can be clamped to optical posts (omitted).}
    \label{fig:RotorMounts_Render}
\end{figure}

There are two main configurations of the rotation mount, the \textbf{reflective mount} intended for reflective optical elements, and the \textbf{transmissive mount} intended for transmissive optical elements. The reflective mount seen in figure \ref{fig:RotorMounts_Render}a has its axis of rotation along the optical element's surface and is suitable for applications involving mirrors and gratings. The transmissive mount, seen in figure \ref{fig:RotorMounts_Render}b, instead has its axis of rotation perpendicular to the surface of the optical element and is more suitable for applications involving polarizers and retarders. Both setups include a selection of holders for different optical elements and applications, like filter revolvers and universal base plates. The setups are made to be mounted on posts, where they can slide and be clamped down in any (square) direction to fit most applications. We provide 3D files with mounts for both metric and imperial post sizes, that is $\varnothing12\,mm$ or $\varnothing0.5\,in.$, as well as the open-source 3D CAD model for further customization.

\subsubsection*{Components}
We divide the components used for the rotation mount into two categories, \textbf{3D printed parts} and \textbf{off-the-shelf components}. Parts that can be 3D-printed include a base mount (the frame) and various rotor wheels used for different applications. They are designed in Rhinoceros 3D [\textit{Rhino 6}, Robert McNeel \& Associates] and the 3D CAD model (.3dm-file) can be found in the resources \cite{bib:Nilsson2020-2, bib:Nilsson2020-3}, along with exported .stl (\textit{stereolithography}) and .stp-files (AP 214, ISO 10303-21). These parts are designed to be made with a \textit{fusion deposition modeling} (FDM) printer, allowing for rapid prototyping and in-house manufacturing.

Our mounts require a rigid and strong material to function properly. Common plastics like \textit{polylactic acid} (PLA), \textit{acrylonitrile butadiene styrene} (ABS) and \textit{polyethylene terephthalate} (PET/PETG) are all suitable alternatives, with the latter providing more durability \cite{bib:hanon2019}. We designed the mounts to print on almost any 3D printer without using support material with a wide range of parameter settings. We recommend a layer height of 0.4 \textit{mm} and nozzle size of 0.6 \textit{mm}, or less. The infill can be as low as 20\%, and to increase the strength the number of perimeters (shells) should be 3 or more. Further, which parts are needed depends on the application. To help select what parts to print, we provide a guide in table \ref{tab:3DPartList} in the appendix. In places where tight tolerances are needed, like thru-holes for shafts and rods, finishing work can be required. If a hole is too tight, it can be honed out with a simple hand drill.

The off-the-shelf components also depend on the configuration and, for this, part lists are provided in appendix \ref{subsec:ComponentList}. Even so, all rotation mounts use a standard NEMA 17 series stepper motor. We have in this work used the cheapest and most common one [\textit{17H2A4413}, MotionKing], with a rotor resolution of $200\,fullsteps/revolution$ ($1.8^{\circ}$) and a holding torque of $40\,N\cdot cm$, which is the recommended minimum. There is an option to use factory-made aluminium pulleys for the transmissive mount, which is what we used in the early stages of development. However, these require extra machining to fit our application and we noted no significant changes when shifting to the 3D printed equivalent. If selected, the lower drive wheel [\textit{21T5/16-2}, RS PRO] needs a spacer for the drive shaft in the form of a $L21x\varnothing_{in}5x\varnothing_{out}6\,mm$ brass sleeve, and the upper rotor wheel [\textit{21T5/32-2}, RS PRO] needs machining work, as described by figure S2 in the supplementary material.

Lastly, we manage to build the rotation mount described here for approximately \texteuro 150, using the components specified in the appendix and assuming access to a 3D printer and basic tools. Note that the price for some parts might increase due to minimum order quantity requirements for some suppliers. As an example, we show the components needed for a transmissive mount in figure S1.

\subsubsection*{Build Guide}
The assembly of a rotation mount starts by fitting the four M6 x 16 \textit{mm} screws and nuts to the base mount (the frame), these are used to clamp it to the mounting posts. After this, the stepper motor is inserted and secured through the front with four M3 x 8 \textit{mm} screws. Now, depending on the configuration the following procedure will differ. For the \textbf{reflective mount}, a rotor base plate is mounted directly to the drive shaft of the stepper motor with an M3 x 6 \textit{mm} set screw and nut. To this, an optic element holder is attached by a clockwise quarter rotation of the pretensioning fork (different rotors/holders are available). The optical element is clamped down from the top by an M4 x 6 \textit{mm} set screw and nut. The position of its front face can be adjusted to align it with the rotation axis and this is done with an M4 x 25 \textit{mm} screw and nut. Two additional nuts are used to lock the screw and remove the slack (a spring can be used instead).

For the \textbf{transmissive mount}, the drive wheel is mounted to the stepper motor with an M3 x 6 \textit{mm} set screw and nut. The drive belt then transfers this rotation to the rotor wheel and this is mounted to the cage rotation mount [\textit{CRM1/M}, Thorlabs] with four No.4 x 6 \textit{mm} screws. With the drive belt wrapped around both wheels, the cage rotation mount can be secured from the sides by four more No.4 x 6 \textit{mm} screws. This should be done to give the drive belt suitable tension and a good starting point is at $4\,mm$ belt movement when applying $0.5\,kg$ of force, as discovered while measuring backlash in section \ref{sec:Test}.\ref{subsec:Criteria} below. Lastly, with all the parts and tools at hand, the construction and incorporation of this rotation mount into an optical system can be done within a day and that includes the building of a control unit (driver), as described next.

\subsection{Building the Controller}
\subsubsection*{Components \& Wiring}

\begin{figure*}[!t]
\centering
\includegraphics[width=0.95\linewidth]{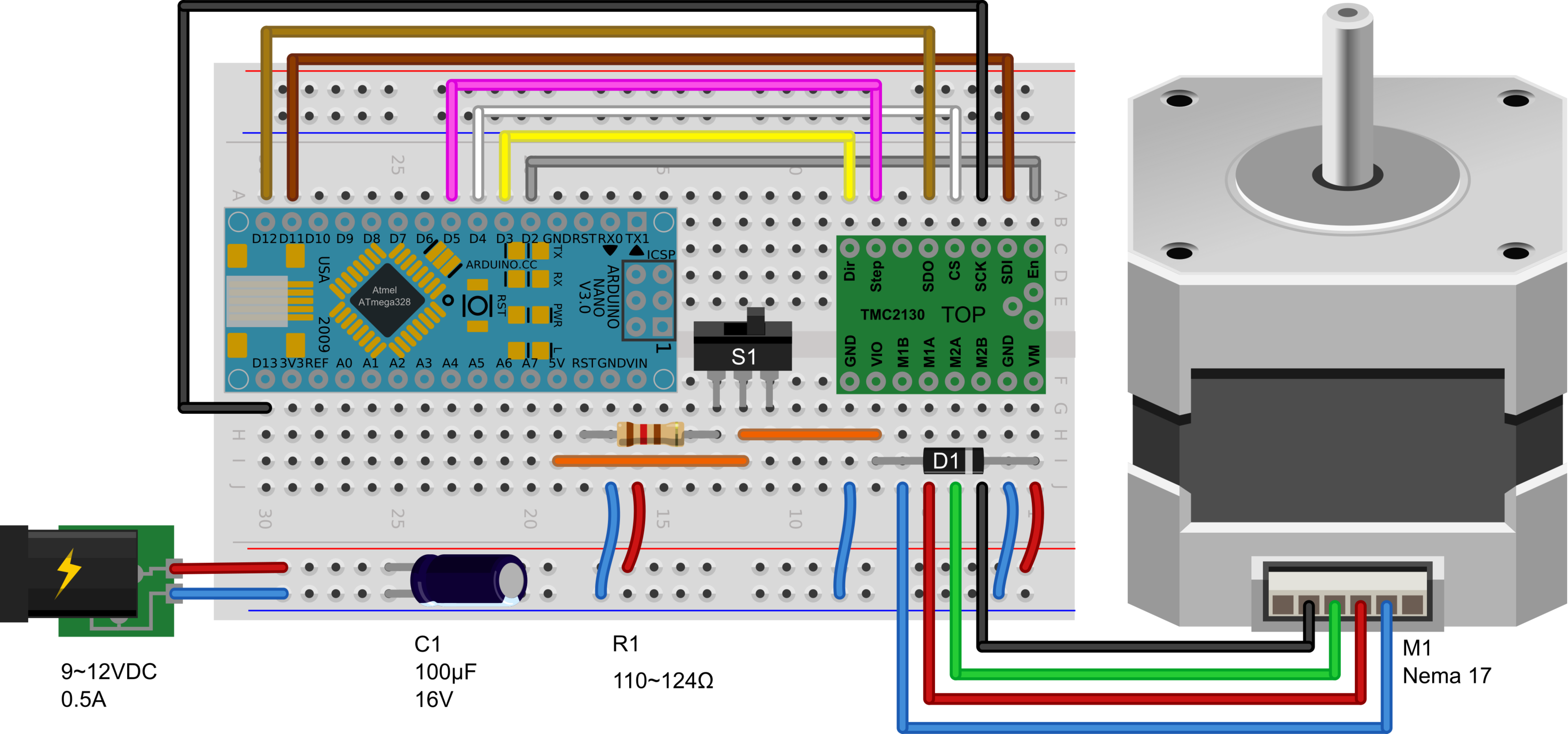}
\caption{The minimum electronics and wiring needed to run a rotation mount when using the MWE code. This setup uses an Arduino Nano microcontroller, a TMC2130 stepper driver and a NEMA 17 stepper motor, all mounted to a half size solderless breadboard.}
\label{fig:Wiring}
\end{figure*}

To control the rotation mounts we designed a controller based on a standard Arduino board \cite{bib:Badamasi2014} and a SilentStepStick stepper motor driver [\textit{TMC2130}, TRINAMIC]. There are many Arduino board models suitable for this controller, and here we show the two most common models that can be used. The first is the \textbf{Arduino Nano}, its small size and PCB compatible headers are perfect for permanent installations and enclosures. The second is the \textbf{Arduino Uno}, its larger size and accessibility is handy for bench testing and experimentation. However, both of these have similar performance and wiring.

The control unit can be powered by a $9 - 12\,VDC$ power adapter and to stabilize the supply, a $100\,\mu F$ (C1) electrolytic capacitor is connected between the supply rail and ground (GND). The supply rail is connected to the raw input of the Arduino (VIN) and the motor input of the driver (VM). The regulated output from the Arduino (5V) is connected to the logical input of the driver (VIO) and a Schottky diode (D1) is connected back to the supply rail, with the cathode at the supply rail (allowing for a safe connection of USB before power supply).

In addition, a switch (S1) or a jumper can be used to toggle the auto-reset mode of the microcontroller on/off (used for uploading code), where a closed connection turns the mode off (this allows the controller to remain running during the loss of USB connection). On some Arduino boards (e.g. Nano V3.0), this can be realized by wiring the switch in series with a $120\,\Omega$ resistor (R1) and connecting them between the reset pin of the Arduino (RST) and the regulated output (5V). However, on other Arduino boards (e.g. Uno R3) there are two solder pads (RESET EN) that can be cut and connected to a switch with the same function.

The rest of the connections are described in figure \ref{fig:Wiring}, along with all the electrical components (see appendix \ref{subsec:ComponentList}) and wiring needed to control a rotation mount. Note that the wire colors of the stepper motor (M1) may vary between manufacturers and that each coil can be found with a basic continuity measurement. We show the wiring for a bi-polar (4 wires) stepper motor, but a uni-polar (6 wires) stepper motor can also be used by connecting the center tap of each of the two coils together. Additionally, the stepper motor driver can be fitted with a small heat flange on top to increase system stability and these are often provided by the manufacturer.

\subsubsection*{Programming}
The controller needs to be programmed before it can be used to control a rotation mount and, for this, we provide a Minimal Working Example (MWE) code called \textbf{RotorMountController\_MWE}. With this code, the angle of a rotation mount can be controlled from a computer and with floating-point precision. A more comprehensive code, called \textbf{RotorMountController\_Multiple} is also provided and both of these are available in the resources \cite{bib:Nilsson2020-4}. The latter shows how to control four different rotation mounts (and drivers) simultaneously using one Arduino Nano/Uno, as well as how to incorporate software emulated origins, backlash compensation, intensity control, and more.

To compile and upload code to the board, we used the open-source Arduino IDE (V.1.8.12). This requires some additional (open-source) libraries to be downloaded and these are described in the code. When proceeding, remember to select the correct rotor resolution, gear ratio, and step size (\textit{microsteps}), these are defined in the code. However, in case of problems to compile the code or download libraries, pre-compiled binary files (.ino.hex) of the MWE code can be found in the resources \cite{bib:Nilsson2020-4} for both Arduino boards and rotation mounts (for a rotor resolution of $200\,fullsteps/revolution$ and a step size of $128\,microsteps$). The procedure for uploading a binary file in Windows is described in detail under section \ref{sec:Appendix}.\ref{subsec:Upload}.

The USB interface of the controller can be accessed directly by the serial monitor in Arduino's IDE or ones like RealTerm. It can also be integrated into larger automation software, such as LabVIEW, because of its command-based interface. The interface for the MWE consists of two functions, one for setting the rotor angle ("SRA\_<DEGREES>") and one for getting the rotor angle ("GRA" $\to$ DEGREES) from the mount. A help function ("HELP") is also included to aid the operator while testing. When using a serial monitor to talk with the controller, it is important to have chosen the correct Baud rate (57600) and EoL character (CR+LF) for the current system.

\section{Test And Comparison}
\label{sec:Test}
\subsection{Measuring Setup}
\begin{figure}[!t]
    \centering
    \includegraphics[width=\linewidth]{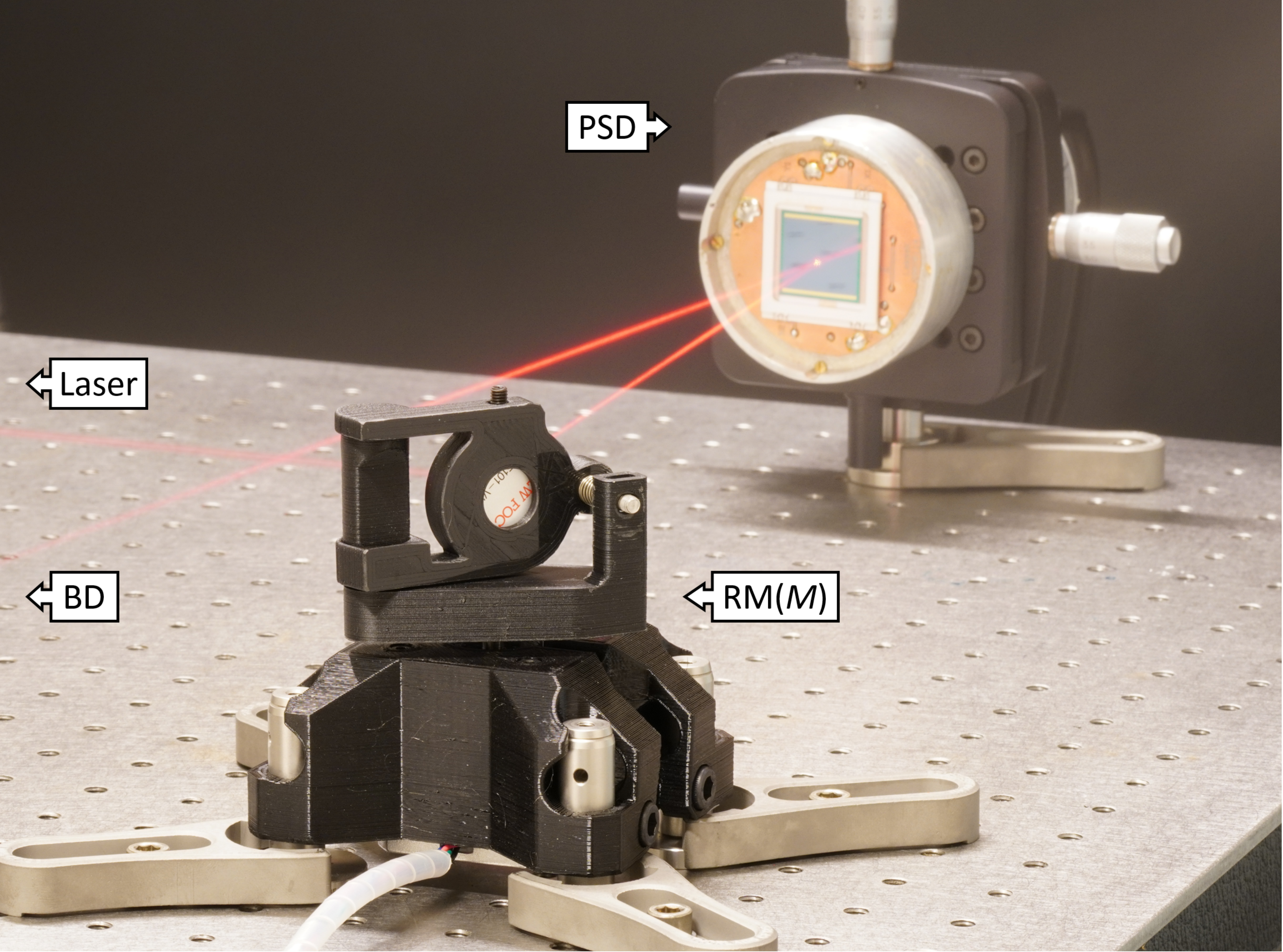}
    \caption{The setup used to test 3D printed rotation mounts, here seen with the reflective mount (RM). A laser \textcolor{red}{beam} is reflected by the rotating mirror (\textit{M}) and its position is measured by the PSD. The remaining light is reflected to a beam dump (BD) and absorbed.}
    \label{fig:Setup}
\end{figure}

Before we can compare our 3D printed rotation mounts to commercially available systems, we need to measure and evaluate their most relevant attributes. To perform these tests, the MWE code is modified by adding single-stepping capabilities and a simple setup is constructed. This setup is used for both the reflective and transmissive mount and can be seen under test conditions in figure \ref{fig:Setup}, except now with ambient light present. The rotation mounts are positioned to rotate in the plane of the table, i.e. with the axis of rotation oriented vertically. Each mount is fitted with a 1" flat mirror [\textit{5101-VIS}, New Focus, Inc.], mounted with its reflective surface along this axis. The rotation of the mounts can then be determined from trigonometry by reflecting a laser beam and measuring its position.

For this, a HeNe-laser ($633\,nm$, $7\,mW$) [\textit{1137/P}, JDS Uniphase] is used and it is reflected towards a $20x20\,mm$ \textit{position sensitive detector} (PSD) [\textit{2L20\_SU9}, SiTek]. This is mounted a given distance $D$ after the rotating mirror, which limits the measurement range of the flat detector plane to $\pm 15\,mrad$ ($2^{\circ}$). The stage rotation is calculated by $\omega = 0.5\cdot\arctan(d/D)$, where the spot distance $d$ on the detector is calibrated from the PSD signal and for one \textit{fullstep}. The signal is collected with a computer, using a \textit{data acquisition} unit (DAQ) [\textit{BNC-2090}, National Instruments]. To avoid aliasing, a low-pass filter [\textit{SR640}, Stanford Research Systems Inc.] is used with a cutoff frequency of $4\,kHz$, while sampling just above the Nyquist frequency ($8192\,Hz$). Data is processed from the DAQ in a custom LabVIEW program [\textit{2018\_V.18.0f2}, National Instruments], see figure S4, and further analyzed in MATLAB\textsuperscript{\textregistered} [\textit{R2020a}, MathWorks]. To aid in measurement accuracy, the setup is built upon a Nexus\textsuperscript{\textregistered} vibration isolated optical table [\textit{T1020CK}, Thorlabs] and situated in a temperature controlled room ($296\pm1\,K$).

\subsection{Evaluation Criteria}
\label{subsec:Criteria}

\begin{figure*}[!t]
    \centering
    \includegraphics[width=\linewidth]{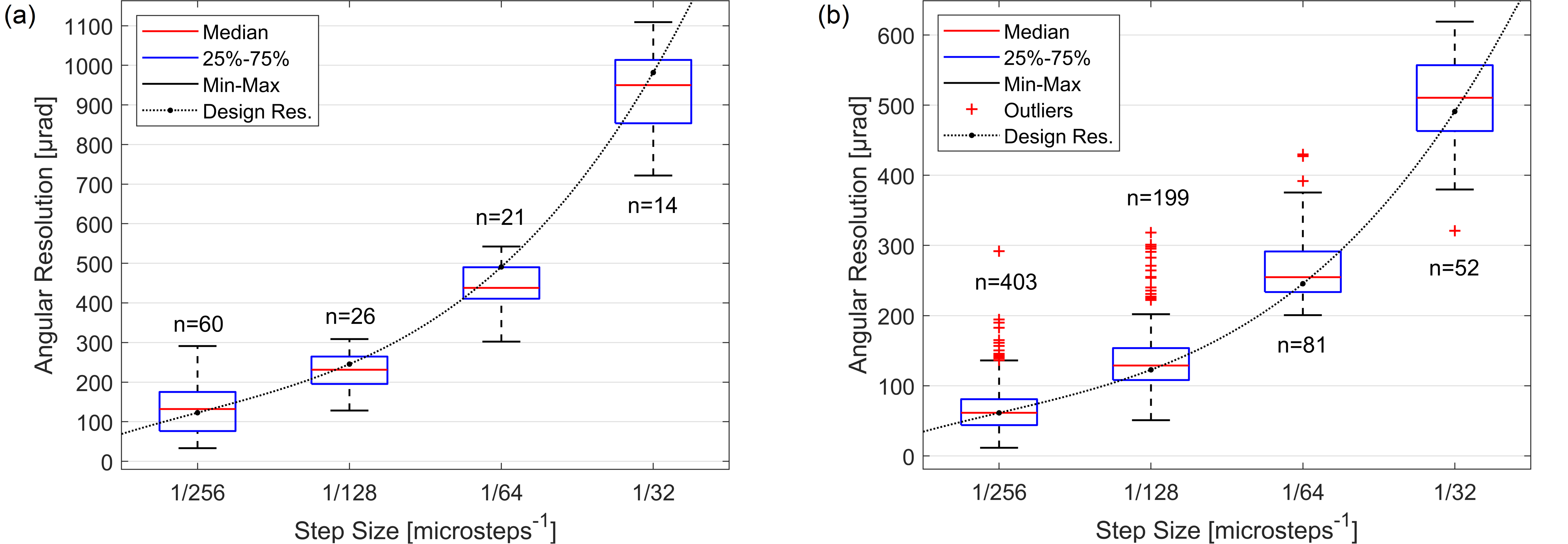}
    \caption{Minimum incremental motion at different step sizes for the reflective mount (a) and transmissive mount (b), calculated from the step response data in figure S5. A stepper motor with $200\,fullsteps/revolution$ ($1.8^{\circ}$) resolution is used and we measure over a stage rotation of $1-2^{\circ}$, resulting in different sample sizes (n) for each set.}
    \label{fig:Resolution}
\end{figure*}

We evaluate the reflective and transmissive mount separately, because of their difference in construction. The results are then compared to manufacturer specifications for commercial mounts with clear apertures of $\varnothing20 - 40\,mm$, see table \ref{tab:Comparison} in the appendix.

\subsubsection*{Velocity}
The rotation of these stages is made in discrete steps, since we are using stepper motors, and the size of these steps can be reduced by dividing them into \textit{microsteps}. A maximum step rate of $4,000\,microsteps/s$ is allowed by the controller when using an Arduino Nano/Uno (clock-rate limited). This is equivalent to $6.4s/rev$ and $12.8s/rev$, for the reflective and transmissive mount, respectively and at a resolution of $128\,microsteps$. If greater velocities are required, the resolution can be lowered by using fewer \textit{microsteps}. The maximum velocity for our setups is reliably achieved at a resolution of $8\,microsteps$, to $0.4s/rev$ and $0.8s/rev$, respectively. This is limited by the strength of our stepper motor but is still higher than that reported for most commercial systems.

\subsubsection*{Resolution}
The minimum incremental motion, here denoted (angular) resolution, is a property dependent on both the stepper motor, the transmission and the controller. It is defined as the change in steady-state angle between consecutive steps and this is to disregard intermediate overshoots, as seen in figure S5a. We start by investigating the \textbf{reflective mount} with its direct transmission (1:1). Since we always use the same stepper motor, the step size of the driver becomes decisive for the final resolution, and this driver can divide each \textit{fullstep} into a much as $256\,microsteps$.

Looking at the median resolution for each step size in figure \ref{fig:Resolution}a, it can be assumed that the experimental data is in good accordance with the theoretical design resolution and this would give a "typical" resolution of $110\pm60\,\mu rad$ at $256\,microsteps$ for the reflective mount. However, we would argue that the maximum value for each set would constitute a better estimate for its nominal resolution. On this basis and while taking into account the decline in stepper motor torque (and hence acceleration) for smaller step sizes, a value of $128\,microsteps$ seems to give an adequate step size and a "guaranteed" resolution of $310\,\mu rad$.

Now, looking at the \textbf{transmissive mount} with its belt transmission (2:1) in figure \ref{fig:Resolution}b. A "typical" resolution of $70\pm30\,\mu rad$ is achieved at $256\,microsteps$, but the departure from the design resolution has become more apparent. It shows an increasing number of outliers at smaller step sizes and these are caused by the delay and (non-returning) overshoot of steps, so-called stick-slip behavior. These can be seen in the raw step response data in figure S5b and are a result of static friction in the cage rotation mount [\textit{CRM1/M}, Thorlabs] and play (clearance) in the belt transmission. Consequently, this will reduce the "guaranteed" resolution and a value of $310\,\mu rad$ is again observed at $128\,microsteps$.

The maximum resolution is on par with many (but not all) commercial stages and this is a compromise done to lower cost and ease manufacturing. With that said, for many applications, this resolution is more than adequate. For example, in an optical attenuator setup, the transmitted intensity $I$ is given by \textit{Malus’s law} \cite{bib:Collett} as
\begin{equation*}
    \label{eq:Malus}
    I = I_0 \cdot \cos^{2}\theta_i \quad,
\end{equation*}
where $I_0$ is the input intensity (or irradiance) of the light and $\theta_i$ is the angle between the light's polarization and the fast axis of the polarizing element. Its derivative describes the final resolution in intensity for the setup and can be written as $I'=-I_0\sin{2\theta}$, which decreases at low intensities ($\theta_i\to\pi/2\,rad$). Thus, the resolution is much better ($I'\to0$) at low intensities, where it is often needed the most, compared to in the middle. Even so, a "guaranteed" resolution of $0.03$ percentage points intensity is achieved in the middle for both rotation mounts and this is usually more than enough.

\subsubsection*{Accuracy}
The uni-directional repeatability is commonly used as a measure for accuracy and it shows how close to a reference position it will come while returning from the same direction and a suitable distance away. Here we measured this by rotating the stages in increments of full revolutions ($2\pi\,rad$) and finding the maximum deviation between these. This results in an accuracy of $250\,\mu rad$ and $600\,\mu rad$, for the reflective and transmissive mount, respectively and at $128\,microsteps$ resolution. An accuracy that falls within that of most commercial mount. Moreover, the absolute position is not defined when using an open-loop system like a stepper motor. So to aid in repeatability, a homing switch is usually incorporated into the setup. For the sake of simplicity, we instead show how to implement SW emulated origins in the program of the controller.

\subsubsection*{Backlash}
The backlash, also called bi-directional repeatability or hysteresis, is a measure of the total accumulated play in a system and is noticeable as an angular loss when altering rotational direction \cite{bib:Merzouki2004}. For the reflective mount (direct drive), there is no significant mechanical play in the system but rather the backlash is a product of the stepper motor torque and bearing friction. For the transmissive mount (belt drive), the backlash is dominated by the transmission and is a product of the belt tension. Removing slack in the belt by increasing its tension will generally decrease the backlash. However, if the belt tension is too high, excess friction in the cage rotation mount [\textit{CRM1/M}, Thorlabs] will arise, resisting the torque of the stepper motor and instead contribute to an increase in backlash.

A compromise between these must therefore be found and for our mount, this is at $4\,mm$ movement when applying $0.5\,kg$ of (normal) force to the belt, midway between the wheels. This results in a backlash of $700\,\mu rad$ and $7000\,\mu rad$, for the reflective and transmissive mount, respectively and at $128\,microsteps$ resolution. Additionally, an alternative technique for backlash reduction can be implemented in software and this is shown in \textbf{RotorMountController\_Multiple}. With this compensation algorithm tuned for the transmissive mount, the backlash can be reduced by a factor of 10 or more and this results in backlash smaller than that for most commercial mounts.

\subsubsection*{Axis Wobble}
Changes in tilt of the rotation axis during stage movements are called axis wobble. We estimate this with the same setup as before, however, we now measured the departure from the horizontal axis. This is done by sampling the vertical position of the spot on the PSD. The results, seen in figure S6, show the axis wobble for a stage rotation of $2^{\circ}$. The $95\%$ confidence bounds give an axis wobble of $\pm52\,\mu rad$ and $\pm35\,\mu rad$, for the reflective and transmissive mount, respectively. This is less wobble than that for most commercial mounts and as expected, the plain bearing in the cage rotation mount (transmissive) gives smoother operation than the ball bearing in the stepper motor (reflective). Note that these measurements were only done for a limited range and are dependent on the quality of the components used.

\subsubsection*{Stability}
Angular stability is crucial for the reflective mount to not introduce disturbances into the optical system. We estimated the stability by measuring the angle of a stationary rotation mount for long periods of time and using Allan deviation calculations \cite{bib:allan1966, bib:Hopcroft2021}. The result is presented in figure S7 and it shows that there is no increase in fluctuations when turning on the stepper motor power as opposed to when it is turned off. The predominant deviations are observed at the highest frequencies ($1\,kHz$) and are most likely originating in electrical interference of the measurement equipment. Still, the angular fluctuations of the mount are in both cases negligible and in orders of magnitude smaller than the resolution, at less than $2\,\mu rad$. Moreover, the low stepper noise of this setup is important while performing sensitive measurements, and this would not be possible using the piezoelectric rotor from previous works. This capability is introduced by StealthChop\textsuperscript{\texttrademark} technology, found in newer stepper motor drivers.

\section{Example Application}
\begin{figure}[!t]
    \centering
    \includegraphics[width=\linewidth]{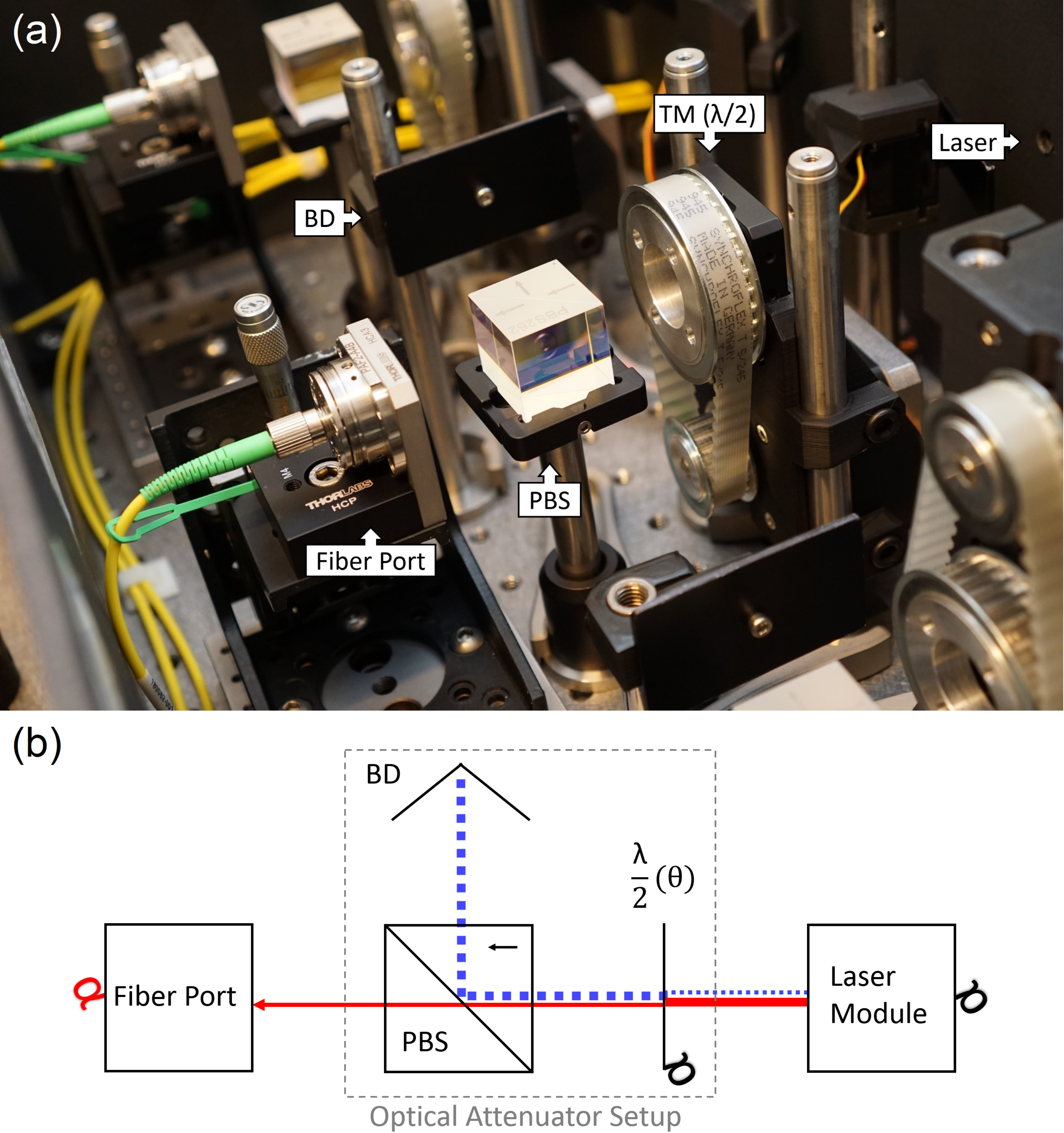}
    \caption{The optical attenuator setup uses a \textbf{transmissive mount} (TM) to rotate a retarder ($\lambda/2$) and change the polarization direction, thus controlling the output power of the laser. This shows an application for the rotation mount, using it to remotely control parameters in an optical system. (a) Multiple attenuator setups each using a TM and one is mounted upside down to allow for a lower beam path. (b) A schematic representation of the optical attenuator setup. The dashed (\textcolor{blue}{blue}) and solid (\textcolor{red}{red}) lines represents the linear polarization components of the laser light, propagating from right to left. The first gets reflected by the \textit{polarizing beam splitter} (PBS) and absorbed by the \textit{beam dump} (BD), while the second gets transmitted towards the optical fiber.}
    \label{fig:Attenuator}
\end{figure}

As a proof of concept, we show how a rotation mount can be used to control the output power of multiple lasers used for optical tweezers and Raman spectroscopy measurements \cite{bib:Stangner2018, bib:Malyshev2021}. To fill the back aperture of our microscope objective and to get a strong lateral optical trap, we need an attenuator that can handle a beam diameter up to $20\,mm$. A detailed description of the system is shown in \cite{bib:Nilsson2020} and the attenuator setup is seen in figure \ref{fig:Attenuator}. An earlier iteration of the transmissive mount is used to rotate a retarder ($\lambda/2$) at the exit of each laser. In conjunction with a \textit{polarizing beam splitter} (PBS), this allows for attenuation of the laser output before coupling it towards the optical tweezers setup (via optical fibers). This gives us control of the trap stiffness and composition without affecting the enclosed system's stability (thermal/mechanical). The transmissive mount is suitable for in-line optical elements and can be further developed for other applications, like spatial filtering (aperture control), etc. The reflective mount came about with the intent to allow for the construction of spectrometers and other scanning applications. However, it can also be used as a beam selector, filter revolver or other application requiring stable and silent rotation \cite{bib:Stangner2017}.

\section{Conclusion}
We show an easy way of constructing rotation mounts using tools and equipment available in most labs. By keeping the design simple and instead implementing functions in the program of the control unit, we have reduced the number of components, the price and the build time considerably in comparison to previous works. A rotation mount and driver can be built within a day for no more than \texteuro 200. Still, their performance is sufficient enough for most applications and on par with commercial systems costing more than \texteuro 2000.

We provide component lists, build instructions and example codes for two types of rotation mounts. We also provide open-source 3D models for the manufacturing and further modifications of the system. A comprehensive list of commercial stages and their specifications are given in table \ref{tab:Comparison}, to aid researchers in need of a motorized rotation mount.







\section{Additional Information}
\subsection*{Funding}
This project is financially supported by the Swedish Foundation for Strategic Research and the Swedish Research Council (2019-04016).



\subsection*{Disclosures}
The authors declare no conflicts of interest.

\section{Appendix}
\label{sec:Appendix}

\subsection{Off-the-shelf Component List}
\label{subsec:ComponentList}
\noindent\textbf{Transmissive Mount:}
\begin{itemize}
\setlength\itemsep{1pt}
    \item 1 pc - NEMA 17 Stepper Motor [\textit{17H2A4413}, MotionKing]
    \item 1 pc - Drive Belt [\textit{10/T5/245SS}, Contitech] (see figure S3)
    \item 1 pc - Cage Rotation Mount [\textit{CRM1/M}, Thorlabs]
    \item 2 pcs - $\varnothing$12 x 150 \textit{mm} Optical Post [\textit{TR150/M-JP}, Thorlabs]
    \item 4 pcs - M6 x 16 \textit{mm} Socket Screw [\textit{HK76168}, Holo-Krome]
    \item 4 pcs - M6 Hex Nut [189-591, RS PRO]
    \item 8 pcs - No.4 x 6 \textit{mm} Socket Screw [\textit{HK72018}, Holo-Krome]
    \item 4 pcs - M3 x 8 \textit{mm} Socket Screw [\textit{HK76012}, Holo-Krome]
    \item 1 pc - M3 x 6 \textit{mm} Set Screw [\textit{529-911}, RS PRO]
    \item 1 pc - M3 Hex Nut [\textit{189-563}, RS PRO]
\end{itemize}
\noindent\textbf{Reflective Mount:}
\begin{itemize}
\setlength\itemsep{1pt}
    \item 1 pc - NEMA 17 Stepper Motor [\textit{17H2A4413}, MotionKing]
    \item 4 pcs - $\varnothing$12 x 50 \textit{mm} Optical Post [\textit{TR50/M-JP}, Thorlabs]
    \item 4 pcs - M6 x 16 \textit{mm} Socket Screw [\textit{HK76168}, Holo-Krome]
    \item 4 pcs - M6 Hex Nut [189-591, RS PRO]
    \item 1 pc - M4 x 25 \textit{mm} Socket Screw [\textit{HK76080}, Holo-Krome]
    \item 1 pc - M4 x 6 \textit{mm} Set Screw [\textit{529-949}, RS PRO]
    \item 4 pcs - M4 Hex Nut [\textit{527-252}, RS PRO]
    \item 4 pcs - M3 x 8 \textit{mm} Socket Screw [\textit{HK76012}, Holo-Krome]
    \item 1 pc - M3 x 6 \textit{mm} Set Screw [\textit{529-911}, RS PRO]
    \item 1 pc - M3 Hex Nut [\textit{189-563}, RS PRO]
\end{itemize}

\noindent\textbf{Control Unit:}
\begin{itemize}
\setlength\itemsep{1pt}
    \item Microcontroller Board [\textit{Nano V3.0} or \textit{Uno R3}, Arduino]
    \item SilentStepStick Driver Board [\textit{TMC2130}, TRINAMIC]
    \item $12\,VDC$ Power Adapter [\textit{GST36E12-P1J}, MEAN WELL]
    \item $100\,\mu F$ Electrolytic Capacitor [\textit{ECA1CM101}, Panasonic]
    \item Schottky Diode [\textit{1N5402RLG}, ON Semiconductor]
    \item 2.1 x 5.5 \textit{mm} Barrel Jack [\textit{RND 205-00905}, RND Connect]
    \item Slide Switch (SPST) [\textit{MFS 131 D}, KNITTER-SWITCH]
    \item $120\,\Omega$ Axial Resistor [\textit{174-2788}, RS PRO]
    \item Add'l: Breadboard, jumper cables, enclosure, USB-cable.
    \end{itemize}

\subsection{Upload Compiled Binary File with Arduino IDE (V.1.8.12) in Windows (7/8/10)}
\label{subsec:Upload}
\begin{enumerate}
\setlength\itemsep{1pt}
    \item Connect the Arduino board to a computer via USB.
    \item Start the Arduino IDE and open the Blink sketch \\(File$\to$Example$\to$1.Basic).
    \item Choose the correct board type and serial port \\(Tools$\to$Board/Port).
    \item Turn on output during uploading (File$\to$Preferences $\to$Settings$\to$Show verbose output during: upload)
    \item Upload the sketch (Sketch$\to$Upload).
    \item In the output panel of the IDE, find and copy the AVR-Dude call, i.e. the first line after the \textit{''Global variables use...''} line.
    \item Paste this command into any text editor and replace the rightmost file path (C:$\backslash$Users$\backslash$...$\backslash$arduino\_build\_XXXXXX/ Blink.ino.hex:i) with the absolute file path to the new binary file (C:$\backslash$Users$\backslash$.../RotorMountController\_MWE.ino.hex).
    \item For Windows 8 or never: Put all three (3) file paths inside of quotation marks (-CC:$\backslash$Users$\backslash$.../avrdude.conf $\Rightarrow$ -C"C:$\backslash$Users$\backslash$.../avrdude.conf", etc.)
    \item Open the Command Prompt (Ctrl+Esc$\to$"cmd"), then paste and run the modified command.
\end{enumerate}

\begin{landscape}
\begin{table}[!h]
\centering\small
\caption{A comprehensive list of commercial motorized rotaton mounts (360°) for $20-40\,mm$ optics, including corresponding controllers and sorted after the total price for one mount and controller. \textbf{DIY} denote the Do-It-Yourself mounts, where RM and TM are the reflective and transmissive mounts and \textit{CCRMLDO} a mount by Rakonjac et al. \cite{bib:Rakonjac2013}\\}
\label{tab:Comparison}
\begin{tabular}{llllllllllll}
\hline
\multicolumn{11}{c}{\textbf{Motorized Rotation Mounts:}} \\
Manufacture: & \textbf{DIY} & \textbf{DIY} & \textbf{DIY} & Standa & Thorlabs & EKSMA Optics & Thorlabs & PI & LK-Instruments & Newport & PI \\
Model: & RM & TM & CCRMLDO & 8MRU & K10CR1 & 960-0161 & PRM1Z8 & DT-80 & M101A & PR50PP & RS-40 \\
Optic Mount: & "Multiple" & SM1 (1"-40) & Ø40 \textit{mm} & 1" (M27x1) & SM1 (1"-40) & 1" (M27x1) & SM1 (1"-40) & Ø40 \textit{mm} & SM1 (1"-40) & NP1 (1"-20) & Ø20 \textit{mm} \\
Motor Type: & Stepper & Stepper & Piezo Rotor & Stepper & Stepper & Stepper & DC Servo & Stepper & Stepper & Stepper & Stepper \\
Gearing: & 1:1 & 2:1 & 1:1 & 3:1 & 120:1 & 180:1 & 252:1 & 180:1 & 3:1 & 63:1 & 90:1 \\
Velocity: \footnotemark[1] & $15.7\,rad/s$ & $7.85\,rad/s$ & $2.7\,rad/s$ & $3.8\,rad/s$ & $0.18\,rad/s$ & $0.87\,rad/s$ & $0.44\,rad/s$ & $0.52\,rad/s$ & $5.2\,rad/s$ & $0.35\,rad/s$ & $0.12\,rad/s$ \\
Torque: & $0.4\,N\cdot m$ & $0.7\,N\cdot m$ & $0.052\,N\cdot m$ & $0.26\,N\cdot m$ & $0.14\,N\cdot m$ & $0.5\,N\cdot m$ & $0.3\,N\cdot m$ & $0.1\,N\cdot m$ & $0.66\,N\cdot m$ & $0.1\,N\cdot m$ & $0.2\,N\cdot m$ \\
Resolution: & $110\pm60\,\mu rad$ & $70\pm30\,\mu rad$ & $90\pm90\,\mu rad$ & $1300\,\mu rad$ & $520\,\mu rad$ & $22\,\mu rad$ & $8.7\,\mu rad$ & $69.8\,\mu rad$ & $160\,\mu rad$ & $350\,\mu rad$ & $87\,\mu rad$ \\
Accuracy: & $250\,\mu rad$ & $600\,\mu rad$ & $520\,\mu rad$ & - & $60\,\mu rad$ & - & $520\,\mu rad$ & $175\,\mu rad$ & $270\,\mu rad$ & $260\,\mu rad$ & $87\,\mu rad$ \\
Backlash: & $700\,\mu rad$ & $700\,\mu rad$ \footnotemark[2] & - & - & $200\,\mu rad$ & - & $5200\,\mu rad$ & $3500\,\mu rad$ & $1700\,\mu rad$ & $1300\,\mu rad$ & $700\,\mu rad$ \\
Axis Wobble: & $100\,\mu rad$ & $70\,\mu rad$ & $13000\,\mu rad$ & $580\,\mu rad$ & $500\,\mu rad$ & $175\,\mu rad$ & $200\,\mu rad$ & $100\,\mu rad$ & - & $50\,\mu rad$ & $35\,\mu rad$ \\
Homing: & SW \footnotemark[3] & SW \footnotemark[3] & Optical & Switch & Hall & Optical & Switch & Switch & Hall & Optical & Hall \\
Unit Price: \footnotemark[4] & $\sim$ 75 € & $\sim$ 150 € & $\sim$ 700 € & 655 € & 1268 € & 687 € & 876 € & 1210 € & 550 € & 1655 € & 1704 € \\
 &  &  &  &  &  &  &  &  &  &  \\ \hline
\multicolumn{11}{c}{\textbf{ Accompanied Controllers:}} \\
Model: & \multicolumn{2}{c}{Arduino Uno + x TMC2130} & ADuC7020 & 8SMC5-USB-x & "Integrated" & 980-1x45 & KDC101 & C-663.12 & SMCx242 & SMC100PP & C-663.12 \\
Channels (x): & \multicolumn{2}{c}{1 - 4} & 2 & 1 - 4 & 1 & 1 - 4 & 1 & 1 & 2, 4 & 1 & 1 \\
Microsteps: & \multicolumn{2}{c}{1 - 256} & 1 - 180 & 1 - 256 & 1 - 2048 & 1 - 256 & - & 1 - 2048 & 1 - 32 & 2 & 1 - 2048 \\
Unit Price: \footnotemark[4] & \multicolumn{2}{c}{35 € - 75 €} & 30 € & 575 € - 2060 € & 0 € & 730 € - 2060 € & 616 € & 728 € & 1500 €, 2000 € & 715 € & 728 € \\ \hline
\end{tabular}
\begin{flushleft}
\footnotesize
    \quad\footnotemark[1] Maximum velocity as permitted by lowering the resolution/\#\textit{microsteps}.\\
    \quad\footnotemark[2] As reduced from $7000\,\mu rad$ with SW compensation, shown in example code \textbf{RotorMountController\_Multiple}.\\
    \quad\footnotemark[3] Software (SW) homing provided in example code \textbf{RotorMountController\_Multiple}.\\
    \quad\footnotemark[4] As of January 2021 (ex VAT).
\end{flushleft}
\end{table}

\begin{table}[!h]
\centering \small
\caption{A selection guide with all the 3D printed components available for a rotation mount. Branching indicates that there are different options and only one is needed.\\}
\label{tab:3DPartList}
\begin{tabular}{|c|c|c|c|c|c|}
\hline
\multicolumn{4}{|c|}{=== \textbf{Reflective Mount} (RM) ===} & \multicolumn{2}{c|}{=== \textbf{Transmissive Mount} (TM) ===} \\ \hline
\multicolumn{2}{|c|}{RM-Base\_MK2(Post12mm).stl} & \multicolumn{2}{c|}{RM-Base\_MK2(Post0.5in).stl} & TM-Base\_MK3.2(Post12mm).stl & TM-Base\_MK3.2(Post0.5in).stl \\ \hline
\multicolumn{4}{|c|}{} & \multicolumn{2}{c|}{TranmissiveMount-DriveWheel.stl} \\ \hline
\multicolumn{2}{|c|}{RM-RotorHolder\_Base.stl} & \multirow{2}{*}{\begin{tabular}[c]{@{}c@{}}RM-Revolver\\ (8x25mm).stl\end{tabular}} & \multirow{2}{*}{\begin{tabular}[c]{@{}c@{}}RM-Univer\\ sal(M4).stl\end{tabular}} & \multirow{2}{*}{TM-RotorWheel\_OpenBarrel.stl} & \multirow{2}{*}{TM-RotorWheel\_Universal(M4).stl} \\ \cline{1-2}
\begin{tabular}[c]{@{}c@{}}RM-RotorHolder\_\\Round(25mm).stl\end{tabular} & \begin{tabular}[c]{@{}c@{}}RM-RotorHolder\_\\Square(25mm).stl\end{tabular} &  &  &  &  \\ \hline
\end{tabular}
\end{table}
\end{landscape}

See \textbf{Supplement 1} for supporting content.
\bibliography{bibliography}

\ifthenelse{\equal{\journalref}{aop}}{%
\section*{Author Biographies}
\begingroup
\setlength\intextsep{0pt}
\begin{minipage}[t][6.3cm][t]{1.0\textwidth} 
  \begin{wrapfigure}{L}{0.25\textwidth}
    \includegraphics[width=0.25\textwidth]{john_smith.eps}
  \end{wrapfigure}
  \noindent
  {\bfseries John Smith} received his BSc (Mathematics) in 2000 from The University of Maryland. His research interests include lasers and optics.
\end{minipage}
\begin{minipage}{1.0\textwidth}
  \begin{wrapfigure}{L}{0.25\textwidth}
    \includegraphics[width=0.25\textwidth]{alice_smith.eps}
  \end{wrapfigure}
  \noindent
  {\bfseries Alice Smith} also received her BSc (Mathematics) in 2000 from The University of Maryland. Her research interests also include lasers and optics.
\end{minipage}
\endgroup
}{}

\end{document}